\documentclass[12pt]{article} 

\usepackage{amssymb} 
\usepackage{amsmath}
\usepackage{amsfonts}

\newcommand{\R}{\mathbb{R}}

\newcommand{\be}{\begin{equation}}
\newcommand{\bea}{\begin{eqnarray}}
\newcommand{\eea}{\end{eqnarray}}

\newcommand{\kt}{\rangle}
\newcommand{\br}{\langle}

\newcommand{\ed}{\end{document}}
\newcommand{\bbr}{\br\!\br}
\newcommand{\kkt}{\kt\!\kt}

\begin{document}

\title{Erratum: Pseudo-Hermiticity for a class of nondiagonalizable Hamiltonians 
[J.~Math.~Phys.~43,~6343~(2002)~]}
\author{Ali Mostafazadeh\thanks{E-mail address: amostafazadeh@ku.edu.tr}\\ \\
Department of Mathematics, Ko\c{c} University,\\
Rumelifeneri Yolu, 34450 Sariyer, Istanbul, Turkey}
\date{ }
\maketitle

Recently, the authors of \cite{2} used the framework provided in \cite{1} to 
re-examine the consequences of pseudo-Hermiticity for the class of block-diagonalizable Hamiltonians introduced in \cite{1}. In doing so, they discovered that Theorem~2 of \cite{1} did not hold, as they could find a counter-example. This theorem must be replaced with the following.
	\begin{itemize}
	\item[] {\bf Theorem~2:} Let $H$ be as in Theorem~1 of Ref.~\cite{1}. Then $H$ is pseudo-Hermitian if and only if it is Hermitian with respect to an inner product $\bbr~~,~~\kkt$ that supports a positive-semidefinite basis \cite{3} including the eigenvectors of $H$. In particular, for every eigenvector $\psi$ of $H$, $\bbr\psi|\psi\kkt\geq 0$; if the corresponding eigenvalue is real and nondefective (algebraic and geometric multiplicities are equal), $\bbr\psi|\psi\kkt>0$; otherwise $\bbr\psi|\psi\kkt=0$.
	\item[] {\bf Proof:} As shown in \cite{1}, pseudo-Hermiticity of $H$ implies that $H$ is Hermitian with respect to the inner product $\bbr~~,~~\kkt_\eta$ with $\eta$ given by Eq.~(15) of
\cite{1} and $\sigma_{\nu_0,a}=1$. It is not difficult to check that indeed the basis vectors $|\psi_n,a,i\kt$, constructed in \cite{1}, have the property that $\bbr\psi_n,a,i|\psi_n,a,i\kkt\geq 0$,
and that $\bbr\psi_n,a,i|\psi_n,a,i\kkt> 0$ only for the cases that $p_{n,a}=1$ and $E_n\in\R$, i.e.,
$|\psi_n,a,i=1\kt$ is an eigenvector of $H$ with a real eigenvalue. Furthermore, by construction, this basis includes all the eigenvectors of $H$. The proof of the converse is the same as the one given in \cite{1}.
	\end{itemize}
It is important to note that having a positive-semidefinite basis does not imply that the inner product 
$\bbr~~,~~\kkt_\eta$ is positive-semidefinite (unless the Hamiltonian is diagonalizable and has a real spectrum in which case both the basis and the inner product $\bbr~~,~~\kkt_\eta$ are positive-definite \cite{4}.) If the Hamiltonian has defective or complex(-conjugate pair(s) of) eigenvalues, there will always be at least two null vectors with negative \cite{3} linear combinations. Unlike positive vectors, linear combinations of nonnegative vectors need not be nonnegative.

\ed